\documentstyle[12pt]{article}

\title{ Non-extremal Stringy Black Hole}

\author{ Kenji Suzuki \\
\\
\it Department of physics, Tokyo Institute of Technology \\
\it Oh-okayama, Meguro, Tokyo 152, Japan \\
\it ks@th.phys.titech.ac.jp
}

\date{}

\begin{document}

\maketitle

\begin{abstract}

We construct a four-dimensional BPS saturated heterotic string solution from 
the Taub-NUT solution. It is a non-extremal black hole solution 
since its Euler number is non-zero. We evaluate its black hole entropy 
semiclassically. We discuss the relation between the black hole entropy 
and the degeneracy of string states. The entropy of our string solution can 
be understood as the microscopic entropy which counts the elementary string 
states without any complications. 
\vspace{1.5cm}


keywords : entropy, BPS, heterotic string, extremal, Euler number

\end{abstract}

\newpage

\section{Introduction}

The Bekenstein-Hawking black hole entropy has been studied 
semiclassically~\cite{Be,BCH,GH,tH1,tH2,STU,SU,Hor}.
This entropy has recently been interpreted microscopically 
in string theory where the black holes can be identified 
with elementary string excitations~\cite{Sen2}. 
It was shown that stringy effects could correct the Bekenstein-Hawking 
formula for the black hole entropy 
in such a way that it correctly reproduces the logarithm of the density of 
elementary string states. 
Strominger and Vafa~\cite{SV} 
have applied this idea to the entropy of the 5-dimensional 
Reissner-Nordstr$\rm \ddot{o}$m-type (RN-type) extremal black holes, 
which carry axion charges and electric charges. As a result, they found 
that the entropy of their black holes 
counts the degeneracy of the BPS soliton bound states. 
Callan and Maldacena~\cite{CM} have also studied the case of near-extremal 
black holes. 

On the other hand, Hawking and Horowitz~\cite{HHR} have argued that the 
extremal RN-type black holes have no entropy 
because of their topology ${\bf R}^1 \times {\bf S}^1 \times {\bf S}^2$.
This point is also discussed recently by Das et.al~\cite{DDR}. 
Gibbons and Kallosh~\cite{GK} have discussed this idea 
with the Euler number. 
According to their arguments, the entropy of a black hole vanishes if the 
Euler number of the black hole is zero in four-dimensions. 
In two dimensions, clearly the Einstein-Hilbert action is 
proportional to the Euler number. 
In contrast, in four-dimensions there is no simple relation between the 
Einstein-Hilbert action and the Euler number. 
However a relation between the entropy and the Euler number of the black 
hole exists when we consider the diagonal metric.
Using this relation the Euler number is found to vanish if the black hole 
is extremal. 
Therefore we call the black hole whose Euler number is zero extremal. 

The entropy of black holes can be interpreted as the statistical entropy 
that counts the degeneracy of the microscopic states 
in string theories. Strominger and Vafa have applied this idea to the 
extremal black holes in their paper~\cite{SV}. 
However the entropies of the extremal black hole solutions are zero 
according to Hawking's argument. 
Hence it is not simple to discuss the interpretation of black hole entropy 
in string theory using the extremal 
RN-black holes. In order to discuss the issue without these complications, 
we need to find the non-extremal black hole solutions. 

Simultaneously, we need to take into account that the quantum 
correction affects the states of the black hole 
solution and disturbs the counting of states if we use the ordinary black 
hole solutions, for instance the Schwarzschild one. 
In contrast, the quantum correction is small for the BPS saturated states. 
The extremal RN black holes are the BPS saturated solution. However these 
solutions may have no entropy as mentioned above.
Therefore in order to discuss the relation between the entropy of black holes 
and the degeneracy of elementary string states, the solutions we need are 
non-extremal and the BPS saturated states. 

In order to find the BPS saturated solutions, 
the four-dimensional affective heterotic string action was investigated by 
Sen~\cite{Sen1,Sen2,HS}. 
These solutions were led from Schwarzschild and Kerr solutions using the 
$O(7,23)$ transformation. 
However they are extremal because their Euler numbers are zero as it will 
be shown later. 
Therefore his solutions have no entropy, and we need to find another 
solution for the issue. 

The purpose of this paper is to derive a new solution, which is BPS 
saturated and non-extremal. 
We construct this solution from the Taub-NUT solution by using Sen's 
argument. Using this solution, we discuss the 
relation between the black hole entropy and the degeneracy of string states. 

The organization of this paper is as follows. 
In section 2 we review the relation between the entropy of black holes and 
the Gauss-Bonnet action which counts the Euler number. 
We consider the relation between the Euler number and the extremal black hole.
From this consideration, we show that Sen's solutions have no entropy 
because of its Euler numbers. 
In section 3 we study the low energy effective action of four-dimensional 
heterotic string and its symmetry. 
We subsequently derive a new solution by using the Sen's argument. 
In section 4 we evaluate the entropy of this solution semiclassically. 
We then discuss the relation between this black hole entropy and the 
degeneracy of string states using the concept of the 
stretched horizon. 
We present our conclusions in section 5. 

\section{Gauss-Bonnet Action and Extremal Black Hole}

In this section we review a relation between the black hole entropy and the 
Euler number. 
This relation leads to the conclusion that the black hole entropy is zero 
if its Euler number vanishes. 
We then consider the extremal U(1) dilaton solution~\cite{GHS} as an example.
We conclude this section by finding that Sen's metric ~\cite{Sen1,Sen2,HS} 
has no entropy because of its Euler number. 

Firstly we note the relation between the Euler number and the entropy for 
the spherically symmetric metric~\cite{GK,LP,KOP}. 
We consider the following metric: 
\begin{equation}
  ds^2 = - e^{2U(r)}dt^2 + e^{-2U(r)}dr^2 + R(r)^2d\Omega^2. \label{met} 
\end{equation}
This metric corresponds to the verbein forms 

\begin{eqnarray*}
  && \hat{\theta}^{0}=e^{U(r)} dt, \quad \hat{\theta}^{1}= e^{-U(r)} dr, \\
  && \hat{\theta}^{2}=R(r) d\theta, \quad 
     \hat{\theta}^{3}=R(r)\sin \theta d\phi. 
\end{eqnarray*}
The entropy is 
\begin{eqnarray*}
 S_{ent} = \beta E - \ln Z
\end{eqnarray*}
where $E$ is the energy of this system. $Z$ is the partition function 
\begin{eqnarray*}
 Z = e^{-I_E},
\end{eqnarray*}
where $I_E$ is the Euclidean on-shell action. It is explicitly given by 
\begin{eqnarray*}
 I_E = \frac{1}{16 \pi}\int_W(-R) 
 + \frac{1}{8\pi}\int_{\partial W}[K] d\Sigma, 
\end{eqnarray*}
where $W$ is the Euclidean manifold defined from the original black hole 
manifold with $t \to it$. 
$\partial W$ is the boundary of $W$. $[K]$ is the extrinsic curvature. 
$d\Sigma$ is 
\begin{eqnarray*}
 d\Sigma = \hat{\theta}^0 \land \hat{\theta}^{2} \land \hat{\theta}^{3}
\end{eqnarray*}
This action has only one boundary, at spatial infinity $r \to \infty$. 
The reason is 
that the event horizon is not present in the Euclidean action. 

On the other hand, $\beta E$ is represented by 
\begin{eqnarray*}
 \beta E = I_{E,h} 
 = \frac{1}{16 \pi}\int_M(-R) + \frac{1}{8\pi}\int_{\partial M}[K] d\Sigma,
\end{eqnarray*}
where $M$ is the original manifold. 
This action has two boundaries, namely, event horizon and spatial infinity 
because the time translation Killing vector $\frac{\partial}{\partial r}$
is zero at horizon if the manifold is non-extremal. Therefore we need 
to take account of the contribution 
to $E$ from the horizon as well as from the infinity.
Therefore the entropy is 
\begin{eqnarray}
 S_{ent} &=& I_{E,h} - I_E \nonumber \\
      &=& - \frac{1}{8\pi}(\int_{\partial W} - \int_{\partial M}) 
           [K] d\Sigma
           = - \frac{1}{8\pi}(\int_{\partial W} - \int_{\partial M}) 
           [K]
         \hat{\theta}^0 \land \hat{\theta}^{2} \land \hat{\theta}^{3}
          \nonumber \\ 
      &=& \frac{1}{8\pi} \frac{1}{\sqrt{g_{rr}}} 
          \frac{\partial}{\partial r}
          (\int_{\partial W} - \int_{\partial M}) 
           \hat{\theta}^0 \land \hat{\theta}^{2} \land \hat{\theta}^{3}
            \nonumber \\
      &=& \frac{1}{2}(\int_{\partial W} - \int_{\partial M}) 
          R(R \partial_r U + 2 \partial_r R) e^{2U} dt. \label{ent}
\end{eqnarray}

Secondly we study the Gauss-Bonnet action with boundary term, which was led 
by Chern~\cite{Ch}. We define the Gauss-Bonnet action 
of the four-dimensional Riemannian manifold $M^{n}$ ($n$=4) : 
\begin{eqnarray}
  S_{GB} = S_{GB}^{vol} + S_{GB}^{boun} = \chi, \label{chi}
\end{eqnarray}
where $\chi$ is Euler number. $S_{GB}^{vol}$ is the volume term 
\begin{eqnarray} 
  S_{GB}^{vol} = \frac{1}{32\pi}\int_M \epsilon_{abcd}R^{ab} \land R^{cd}, 
 \label{vol}
\end{eqnarray}
and $S_{GB}^{boun}$ is the boundary term 
\begin{eqnarray}
 && S_{GB}^{boun} = -\frac{1}{32\pi^2}\int_{\partial M}
              \epsilon_{abcd}(2\theta^{ab}\land R^{cd}
             + \frac{4}{3}\theta^{ab}\land \theta^c_e \land \theta^{ed}), 
 \label{bound}
\end{eqnarray}
where $\partial M$ is the boundary of $M$. 
In these equations the curvature two-form is defined as 
\begin{eqnarray*}
 R^a_b \equiv d\omega^a_b + \omega^a_c \land \omega^c_b, 
\end{eqnarray*}
where $\omega^a_b$ is a spin connection one-form.
The second fundamental form of the boundary is 
\begin{eqnarray*}
 \theta^{ab} \equiv \omega^{ab}-(\omega^{ab})_0, 
\end{eqnarray*}
where $(\omega^{ab})_0$ is spin connections at the boundary $r=r_0$. 


For the spherically symmetric metric(\ref{met}) 
(which was discussed by Gibbons and Kallosh~\cite{GK,LP,KOP}), 
the spin connections are 
\begin{eqnarray}
  && \omega^{01} = \frac{1}{2} \partial_r (e^{2U})dt, \quad 
     \omega^{21} = e^U \partial_r R d\theta, \nonumber \\
  && \omega^{31} = e^U \partial_r R \sin \theta d\phi, \quad 
     \omega^{32} = \cos \theta d\phi, \nonumber \\
  && (else =0). 
\end{eqnarray}
For this metric, the range of integration in $t$ is infinite in 
the Gauss-Bonnet action (\ref{chi}). 
However if we use the Riemannian version of the metric (\ref{met}), 
\begin{equation}
  ds^2 = e^{2U(r)}d\tau^2 + e^{-2U(r)}dr^2 + R(r)^2d\Omega^2, 
\end{equation}
where $\tau$ is a periodic coordinate, the range of $\tau$ integration is 
constrained to be from $0$ to $\beta$ 
by the standard requirements where 
\begin{eqnarray*}
  \beta = \frac{2\pi}{\kappa}. 
\end{eqnarray*}
In this equation, the surface gravity $\kappa$ is given by 
\begin{eqnarray}
  \kappa &\equiv& \frac{1}{2}\frac{\partial _r g_{tt}}
               {\sqrt{-g_{rr}g_{tt}}}\bigg|_{r = r_H} \nonumber \\
         &=& (\omega^{01})_t  \nonumber \\
         &=& \frac{\partial_r (e^{2U})}{2}\bigg|_{r = r_H},
\end{eqnarray}
where $r_H$ is a radius of the black hole event horizon, which satisfies that 

\begin{eqnarray*}
 (e^{2U})\bigg|_{r = r_H} = 0
\end{eqnarray*}

The Gauss-Bonnet volume integral can be calculated using the values of 
the Riemann tensors for the metric (\ref{met}). 
As a result, the Gauss-Bonnet integrand in the volume term is a total 
derivative. 
The part of the volume term (\ref{vol}) which does not contain 
$\theta^{01}$ is completely cancelled by the boundary term(\ref{bound}). 
Then, the Gauss-Bonnet action for the Riemannian version of this metric is 
\begin{eqnarray} 
  S_{GB} &=& S^{vol}_{GB} + S^{boun}_{GB} \nonumber \\
         &=& \frac{1}{4\pi^2}(\int_{\partial V} - \int_{\partial M})
         \, \omega^{01} \land R^{23} 
               \nonumber \\ 
         &=& \frac{1}{2\pi}(\int_{\partial V} - \int_{\partial M}) 
           \, \partial_r (e^{2U}) (1 - e^{2U}(\partial_r R)^2) dt,
 \label{diag}
\end{eqnarray}
where $M^{2n-1}$ is a $(2n-1)$-dimensional manifold defined from an 
original $n$-dimensional manifold $M^n$ 
whose extra dimensions are formed by the unit tangent vectors. $V$ is an 
$n$-dimensional submanifold of $M^{2n-1}$. 
Chern has shown that the integrand in the Gauss-Bonnet action 
is equal to the exterior derivative of a differential form $\Phi$ in 
$M^{2n-1}$. According to Stokes' theorem, the volume integral is equal to 
the integral over the boundaries of $V$. 
The boundaries of $V$ correspond to the non-isolated singular points of the 
tangent vector field defined from $M^n$. For example, the Schwarzschild 
manifold has two boundaries at its horizon and infinity. The manifold $V$ 
defined from the original manifold has only a single 
boundary at infinity because the horizon of the manifold 
corresponds to the isolated singular point of $V$. 

This expression provides an exact cancellation when the first term and the 
second term are the same, namely the boundaries of the manifold $M^n$ 
are the same as the boundaries of the submanifold $V$ of the manifold 
$M^{2n-1}$. In this case 
the Euler number is zero. 

The boundary $\partial V$ is the same as $\partial W$ defined in (\ref{ent}). 
So we find the relation between the black hole entropy (\ref{ent}) 
and the Euler number (\ref{diag}) :

\begin{eqnarray*}
S_{ent} &=& 2 \pi \chi (e^{2U})'^{-1} (1-e^{2U}R'^2)^{-1}
     \frac{R}{2} \left. [(U'R+2R')]e^{2U} \right|_{r=r_H} \\
 &=& \pi \chi [(e^{2U})'-R'^2 e^{2U}(e^{2U})']^{-1} 
      R[ \frac{R}{2}(e^{2U})'+2R'e^{2U} ]\bigg|_{r=r_H}
\end{eqnarray*}
By defining $r_h$ in such a way that $\left. e^{2U}\right|_{r=r_H}=0$, 
we find :

\begin{eqnarray*}
S_{ent} &=& \frac{\pi \chi R^{2}(r_H)}{2}
        =\frac{\chi (4 \pi R^{2}(r_H))}{8}=\frac{\chi A}{8} \\
  && A =4 \pi R^2 (r_H),
\end{eqnarray*}
Therefore the entropy of the black hole is zero if the Euler number 
of the black hole vanishes. 

Thirdly we define the extremal black holes in an ordinary way. 
We define the temperature of the black hole $T$ as,
\begin{eqnarray}
  T &\equiv& 2\pi \kappa \nonumber \\
    &=& \pi \partial_r (e^{2U})\bigg|_{r = r_H}. 
\end{eqnarray}
Using this quantity, the extremal black hole is defined by 
\begin{equation}
  T = 0. \label{ext}
\end{equation}

As an example, we consider the extremal U(1) dilaton solution~\cite{GHS} : 
\begin{eqnarray}
  ds^2 = - ( 1 - \frac{r_+}{r} )^{\frac{2}{1+a^2}}dt^2 
         + ( 1 - \frac{r_+}{r} )^{-\frac{2}{1+a^2}}dr^2 
         + r^2(1 - \frac{r_+}{r})^{\frac{2a^2}{1+a^2}} d \Omega^2 \\
          (0 \le a \le 1) \nonumber 
\end{eqnarray}
From the above definition, the surface gravity is 
\begin{eqnarray}
  \kappa &=& \frac{1}{2}\partial_r  
             [( 1 - \frac{r_+}{r})^{\frac{2}{1+a^2}}] 
             \bigg|_{r=r_H^+} \nonumber \\
         &=& \frac{2}{1+a^2}(1 - \frac{r_+}{r})^{\frac{1-a^2}{1+a^2}}
             \frac{r_+}{r^2} \bigg|_{r=r_+} \nonumber \\
         &=& \left\{
              \begin{array}{@{\,}ll}
            0  & \mbox{($0 \le a < 1$)}  \\
            \frac{1}{2r_+} & \mbox{($a \to 1$)}. 
              \end{array}
               \right. 
\end{eqnarray}
Then for $0 \le a < 1$, this solution is extremal from the 
definition (\ref{ext}). 
For these solutions, the Gauss-Bonnet actions (\ref{diag}) are 
evaluated to be 
\begin{equation}
  S_{GB} = \frac{1}{4\pi^2}(\int_{\partial V} - \int_{\partial M})
          \omega^{01} (1 - (\frac{a^2}{1+a^2})^2) d\Omega, 
\end{equation}
where 
\begin{eqnarray}
 \frac{1}{2\pi} \int \omega^{01} \bigg|_{r_H} &&= \frac{1}{2\pi}
  \int^{\beta}_{0} dt (\omega^{01})_t \bigg|_{r_H} \nonumber \\
 && = \frac{1}{4\pi}\beta (e^{2U})' \bigg|_{r_H} \\
 && = 1,
\end{eqnarray}
and $r_h$ is the horizon radius of $V$ (or $M$). 
From the knowledge of the topology, the Euler number should be an integer. 
Then in the case $0 < a \le 1$, we are forced to conclude that 
$\partial V = \partial M$, and their Euler numbers are zero. 
In the case $a=0$ (which is the extremal R-N black hole) it 
seems that there is no problem to find $\chi=2$. 
However in $r-\tau$ space ($\tau$ is equal to the imaginary time 
coordinate $it$) the vector field 
$\frac{\partial}{\partial \tau}$ has no fixed points because the surface 
gravity is zero. Then the topology of this space is 
${\bf R}^1 \times {\bf S}^1$, and its Euler number is also zero.
Therefore we find that the Euler numbers are zero for $0 \le a \le 1$. 
From these considerations, we conclude that the extremal black holes are 
defined by : 
\begin{equation}
 T = 0 
\end{equation}
and (or)   
\begin{equation}
 S_{GB} = \chi = 0, \quad (\partial V = \partial M)
\end{equation}

Finally we consider the Sen's solution~\cite{Sen2} : 
\begin{eqnarray}
 &&  ds^2_E = - \Delta^{-1/2}r dt^2 
           + \Delta^{1/2}r{^-1} dr^2
           + \Delta^{1/2}r (d\theta^2 + \sin^2 \theta d\phi^2) \nonumber \\
 &&  \Delta = (r^2 + 2m_0 r \cosh \alpha + m_0^2) 
\end{eqnarray} 
We can calculate the Euler number of this solution using the formula 
(\ref{diag}). The result is that  
\begin{equation}
 S_{GB} =  \frac{1}{4\pi^2}(\int_{\partial V} - \int_{\partial M}) 
            \frac{3}{4} \omega^{01} d\Omega.
\end{equation}
From this formula, we conclude that Sen's solution is extremal since its 
Euler number vanishes.

\section{Heterotic string and Symmetry}

In this section we derive a new solution for the heterotic string action. 
This solution is non-extremal and could possess 
non zero entropy. 

We recall the heterotic string theory that is compactified on a six- dimensional torus~\cite{Sen1,Sen2,HS}. 
The bosonic part of the effective field theory is given by 
\begin{eqnarray}
  S = \frac{1}{32\pi}\int d^4x\sqrt{-G}
      e^{-\Phi} [R_G + G^{\mu\nu}\partial_\mu \Phi \partial_\nu \Phi
    - \frac{1}{12}G^{\mu\lambda}G^{\nu\kappa}G^{\rho\sigma}
      H_{\mu\nu\rho}H_{\lambda\kappa\sigma} 
           \nonumber \\
    - G^{\mu\lambda}G^{\nu\kappa}
      F^{(a)}_{\mu\nu}(LML)_{ab}F^{(b)}_{\lambda\kappa}
    + \frac{1}{8}G^{\mu\nu}Tr(\partial_\mu ML \partial_\nu ML)]. \quad 
   \label{action1}
\end{eqnarray}
where 
\begin{eqnarray}
  && F^{(a)}_{\mu\nu} = \partial_\mu A^{(a)}_\nu - \partial_\nu A^{(a)}_\mu, 
    \nonumber \\
  && H_{\mu\nu\rho} = (\partial_\mu B_{\nu\rho} 
  - 2 A^{(a)}_\mu L_{ab}F^{(b)}_{\nu\rho}) + (\mu,\nu,\rho: cyclic). 
\end{eqnarray}
The massless fields we consider are gravitational fields $G_{\mu\nu}$, 
the anti- symmetric tensor fields $B_{\mu\nu}$, 
twenty-eight U(1) gauge fields $A^{(a)}_{\mu} (1 \le a \le 28)$, the 
scalar dilaton field $\Phi$, and a 
$28 \times 28$ matrix valued scalar field $M$ satisfying, 
\begin{eqnarray}
  MLM^T = L, \quad 
  L = \left(
  \begin{array}{cc}
   -I_{22} & 0 \\
    0 & I_{6}
  \end{array} 
  \right). 
\end{eqnarray}
This action (\ref{action1}) is invariant under the $O(6,22)$ transformation, 
\begin{eqnarray}
  M \to \Omega M \Omega^T, \quad 
  A^{(a)}_\mu \to \Omega_{ab}A^{(b)}_\mu, \quad 
  \Phi \to \Phi, \quad
  G_{\mu\nu} \to G_{\mu\nu}, \quad B_{\mu\nu} \to B_{\mu\nu}.
\end{eqnarray}
where $\Omega$ is a $28 \times 28$ matrix satisfying
\begin{equation}
  \Omega L \Omega^T = L. 
\end{equation}

In addition to that, the action (\ref{action1}) is expected to have an 
$O(7,23)$ symmetry if backgrounds are independent 
of the time coordinate $t$. To see how this appears, we define new 
variables as follows :
\[ 
  \bar{M} = \left(
   \begin{array}{cccc}
    M + 4(G_{tt})^{-1}A_t A_t^T
     & -2(G_{tt})^{-1}A_t 
     & 2MLA_t  \\
      & & + 4(G_{tt})^{-1}A_t(A^T_tLA_t)   \\
      & & \\
    -2(G_{tt})^{-1}A_t^T & (G_{tt})^{-1} & -2(G_{tt})^{-1}A_t^TLA_t \\ 
      & & \\
    2A_t^TLM 
     & -2(G_{tt})^{-1}A_t^TLA_t 
     & G_{tt} + 4A^T_tLMLA_t    \\ 
    + 4(G_{tt})^{-1}A_t^T(A^T_tLA_t) & & + 4(G_{tt})^{-1}(A^T_tLA_t)^2 
   \end{array}
  \right)
\] 
\begin{eqnarray}
   && \bar{A}^{(a)}_i = A^{(a)}_i - (G_{tt})^{-1}G_{ti}A^{(a)}_t, \quad
      \bar{A}^{(29)}_i = \frac{1}{2}(G_{tt})^{-1}G_{ti}, 
       \nonumber \\
   && \bar{A}^{(30)}_i = \frac{1}{2} B_{ti} + A^{(a)}_t L_{ab} A^{(b)}_t, 
      \quad \bar{G}_{ij} = G_{ij} - (G_{tt})^{-1}G_{ti}G_{tj}, 
       \nonumber \\
   && \bar{B}_{ij} 
     = B_{ij}  + (G_{tt})^{-1}(G_{ti}A^{(a)}_j-_{tj}A^{(a)}_i)L_{ab}A^b_t
     + \frac{1}{2}(G_{tt})^{-1}(B_{ti}G_{tj} - B_{tj}G_{ti}), 
       \nonumber \\ 
   && \bar{\Phi} = \Phi - \frac{1}{2}\ln(-G_{tt}), \quad
      \bar{L} = \left(
   \begin{array}{ccc}
    L & 0 & 0 \\
    0 & 0 & 1 \\
    0 & 1 & 0
   \end{array}
    \right), \quad (1 \le a \le 28, 1 \le i \le 3). \nonumber \\
      \label{trans}
\end{eqnarray}  
Using these variables, the action (\ref{action1}) is rewritten as
\begin{eqnarray}
  S = \frac{1}{32\pi} \int dt d^3x
      \sqrt{ \bar{G} }e^{-\bar{\Phi}}[R_{\bar{G}} 
    + \bar{G}^{ij}\partial_i \bar{\Phi} \partial_j \bar{\Phi}
    - \frac{1}{12}\bar{G}^{il}\bar{G}^{jm}\bar{G}^{kn}
      \bar{H}_{ijk}\bar{H}_{lmn}  
       \nonumber \\
    - \bar{G}^{il}\bar{G}^{jm}
      \bar{F}^{(\bar{a})}_{ij}(\bar{L}\bar{M}\bar{L})_{\bar{a}\bar{b}}
      \bar{F}^{(\bar{b})}_{lm}
    + \frac{1}{8}\bar{G}^{ij}
     Tr(\partial_i \bar{M}\bar{L} \partial_j \bar{M}\bar{L})], \quad
    \label{action2} 
\end{eqnarray} 
where
\begin{eqnarray}
  && \bar{F}^{(\bar{a})}_{ij} 
    = \partial_i \bar{A}^{(\bar{a})}_j - \partial_j \bar{A}^{(\bar{a})}_i,
        \nonumber  \\ 
  && \bar{H}_{ijk} = (\partial_i \bar{B}_{jk} 
    - 2 \bar{A}^{(\bar{a})}_i \bar{L}_{\bar{a}\bar{b}}\bar{F}^{(\bar{b})}_{jk})           + (i,j,k: cyclic). 
\end{eqnarray} 
It is obvious that this action (\ref{action2}) has the $O(7,23)$ symmetry :
\begin{eqnarray}
  \bar{M} \to \bar{\Omega} \bar{M} \bar{\Omega}^T, \quad 
  \bar{A}^{(\bar{a})}_i \to 
  \bar{\Omega}_{\bar{a}\bar{b}}\bar{A}^{(\bar{b})}_i, \quad 
  \bar{\Phi} \to \bar{\Phi}, 
  \bar{G}_{ij} \to \bar{G}_{ij},
  \quad \bar{B}_{ij} \to \bar{B}_{ij}, 
\end{eqnarray}
where $\bar{\Omega}$ is a $30 \times 30$ matrix satisfying 
\begin{equation} 
  \bar{\Omega} \bar{L} \bar{\Omega}^T = \bar{L}. 
\end{equation}
In order to simplify $\bar{\Omega}$, we use the orthogonal matrix U that 
diagonalizes $\bar{L}$,  
\begin{equation}
  U = \left(
  \begin{array}{ccc}
    I_{28} & 0                  & 0 \\
     0     & \frac{1}{\sqrt{2}} & \frac{1}{\sqrt{2}} \\
     0     & \frac{1}{\sqrt{2}} & - \frac{1}{\sqrt{2}} 
  \end{array}
  \right). \\
\end{equation}
Then 
\begin{equation}
  U\bar{L}U^T \equiv \bar{L}_d = 
   \left(
   \begin{array}{cccc}
    -I_{22} & 0   & 0   & 0 \\
       0    & I_6 & 0   & 0 \\
       0    & 0   & 1   & 0 \\
       0    & 0   & 0   & -1 
  \end{array}
  \right). 
\end{equation}

From the elements of $L_d$, we see that $\bar{\Omega}$ is equal to $O(6,1) 
\times O(22,1)$, which are the subgroups of $O(7,23)$. In other word, 
the $O(6,1)$ subgroup acts on the 23rd-28th and 30th index, 
whereas the $O(22,1)$ subgroup 
acts on the 1st-22nd, and 29th index. Then we can write 
\begin{equation}
 U \bar{\Omega}_1 U^T = \left(
   \begin{array}{cccccc}
     I_{21} & 0             & 0   & 0           & 0            & 0 \\
       0    & \cosh \alpha  & 0   & 0           & \sinh \alpha & 0 \\
       0    & 0             & I_5 & 0           & 0            & 0 \\
       0    & 0             & 0   & \cosh \beta & 0    & \sinh \beta \\ 
       0    & \sinh \alpha  & 0   & 0           & \cosh \alpha & 0 \\
       0    & 0             & 0   & \sinh \beta & 0    & \cosh \beta \\
   \end{array}
   \right). \\
\end{equation}
In the action (\ref{action2}) we consider the solution with the following 
asymptotic flat space time forms for various fields :
\begin{equation}
  M_{as} = I_{28}, \quad \Phi_{as} = 0, \quad (A^{(a)}_\mu)_{as} = 0, 
  \quad (G_{\mu\nu})_{as} = \eta_{\mu\nu}, \quad (B_{\mu\nu})_{as} = 0. 
  \label{asy}
\end{equation}
Using these fields, 
\begin{equation}
   \bar{M}_{as} = \left(
      \begin{array}{ccc}
        I_{28} & 0  & 0 \\
           0   & -1 & 0 \\
           0   & 0  & -1 
      \end{array}
   \right).
\end{equation}
With these asymptotic fields the action (\ref{action2}) is invariant under 
an $O(6) \times O(22)$ transformation, 
and $\bar{\Omega}$ is parametrized by the coset 
\begin{eqnarray}
 (O(6,1) \times O(22,1)) / (O(6) \times O(22)). \nonumber
\end{eqnarray}
It is because the whole $\bar{\Omega}$ is written as 
$\bar{\Omega} = \bar{\Omega}_2 \bar{\Omega}_1$,
where $\bar{\Omega}_2$ is 
\begin{equation}
   \bar{\Omega}_2 = \left(
   \begin{array}{ccc}
    R_{22}(\vec{n}) & 0 & 0 \\ 
      0  & R_6(\vec{p}) & 0 \\ 
      0  &   0    & I_2
   \end{array}
   \right). \label{omega2}
\end{equation}
In equation (\ref{omega2}) $\vec{n}, \vec{p}$ are arbitrary 22- and 
six-dimensional unit vectors, 
and $R_{N}(\vec{k})$ denotes an $N$-dimensional rotation matrix that 
rotates an $N$-dimensional column 
vector with only the $N$-th component non-zero and equal to 1 into an 
arbitrary $N$-dimensional unit vector $\vec{k}$. 

We can derive the new solutions of the action (\ref{action1}) with the 
following method:

1) Firstly we consider the time-independent solution of general relativity 
such as Kerr or Schwarzschild solutions for $G_{\mu\nu}$, and other fields 
$M, \Phi, A^{a}_{\mu}, B_{\mu\nu}$ are the same as (\ref{asy}).

2) Secondly we apply the transformation (\ref{trans}) to the above solution. 
As a result, we obtain a new solution of the action (\ref{action2}). 

3) Thirdly we apply the transformation $\bar{\Omega}$ to 2). 
This solution is also the solution of the action (\ref{action2}). 

4) Finally we apply the inverse of the transformation (\ref{trans}) to 3), 
and then we get a new solution of the action (\ref{action1}). 

We use the Taub-NUT solution, as the solution 1): 
\begin{eqnarray}
    ds^2 &\equiv& G_{\mu\nu}dx^{\mu}dx^{\nu} 
       = - \frac{r-n}{r+n}(dt + 2in \cos \theta d\phi)^2 \nonumber \\
      && + (r^2-n^2)(d\theta^2 + \sin ^2 \theta d\phi^2) 
         + \frac{r+n}{r-n}dr^2  
\end{eqnarray} 
The Riemannian version of this metric is self dual.
As the result of above transformations 2),3),4), we obtain the following 
solution :
\begin{eqnarray}
  ds^2 & \equiv & e^{-\Phi} G_{\mu\nu}dx^{\mu}dx^{\nu} 
      = - \frac{r-n}{\sqrt{\Delta}}(dt 
      + in \cos \theta(\cosh \alpha + \cosh \beta) d\phi)^2 
           \nonumber \\
      && + (r-n)\sqrt{\Delta}(d\theta^2 + \sin ^2 \theta d\phi^2) 
         + \frac{\sqrt{\Delta}}{r-n} dr^2 
\end{eqnarray}
where
\begin{eqnarray}
  \Delta &=& (r+n)^2 + 2n(r+n)(\cosh \alpha \cosh \beta  -1) 
         + n^2 (\cosh \alpha - \cosh \beta)^2 \nonumber \\
  A_t^{(a)} &=& -\frac{n^{(a)}}{\sqrt{2} \Delta} n \sinh \alpha
         [(r+n) \cosh \beta + n(\cosh \alpha - \cosh \beta)] \quad
       (1 \le a \le 22) \nonumber \\
  A_t^{(a)} &=& -\frac{p^{(a-22)}}{\sqrt{2} \Delta} n \sinh \beta
         [(r+n) \cosh \alpha + n(\cosh \beta - \cosh \alpha)] \quad
       (23 \le a) \nonumber \\
\end{eqnarray}
This solution is not included in ~\cite{CY} because it has Taub-NUT charge.

From the above solution, the ADM mass $m_0$, the surface gravity 
$\kappa$, electric charge $Q^{(a)}$ are found respectively as,
\begin{eqnarray*}
 m_0 = n (1+\cosh \alpha \cosh \beta), \quad 
 \kappa = \frac{1}{2n(\cosh \alpha + \cosh \beta)}, 
\end{eqnarray*}
\begin{eqnarray}
 Q^{(a)} = \left\{
     \begin{array}{@{\,}ll}
     \frac{n}{\sqrt{2}} \sinh \alpha \cosh \beta n^{(a)} 
        & \mbox{($1 \le a \le 22$)}  \\ 
     \frac{n}{\sqrt{2}} \sinh \beta \cosh \alpha p^{(a-22)}  
        & \mbox{($22 \le a$)}. 
     \end{array}
       \right. 
\end{eqnarray}
We define that 
\begin{eqnarray} 
  Q^{(a)}_L = \frac{1}{2}(I_{28}-L)_{ab}Q^{(b)}, \quad 
  Q^{(a)}_R = \frac{1}{2}(I_{28}+L)_{ab}Q^{(b)}. 
\end{eqnarray}
In order to obtain the BPS saturated state, we consider the limit 

\begin{equation} 
  n \to 0, \quad \cosh \alpha \to \infty, \quad n \cosh \alpha =n_0 : 
  {\rm fix}.
\end{equation}
Then the solution takes the form : 
\begin{eqnarray}
 ds^2 &\equiv& G_{\mu\nu}dx^{\mu}dx^{\nu} 
       = - \frac{r}{\sqrt{\Delta'}}(dt + i m_0 \cos  \theta d\phi)^2 
       \nonumber \\ 
      &&+ r\sqrt{\Delta'}(d\theta^2 + \sin ^2 \theta d\phi^2) 
        + \frac{\sqrt{\Delta'}}{r} dr^2,  \nonumber \\
 \Delta' &=& r^2 + 2n_0 r \cosh \beta + n_0^2, \quad
 m_0 = n_0 \cosh \beta, \quad
 \kappa =  \frac{1}{2n_0}, \nonumber \\
 Q_L^{(a)} &=& \frac{n_0}{\sqrt{2}} \sinh \beta n^{(a)} 
            \quad (1 \le a \le 22), \nonumber \\
 Q_R^{(a)} &=& \frac{n_0}{\sqrt{2}} \cosh \beta p^{(a-22)} 
            \quad (22 \le a), \nonumber \\
           && (else = 0). \nonumber 
\end{eqnarray}
From this solution we get the following relations between $m_0, Q_L, Q_R$ :
\begin{equation}
   m_0^2 = 2 \vec{Q}_R^2, \quad m_0 = \sqrt{2} |\vec{Q}_L| \tanh \beta,
    \quad  n_0 = \sqrt{m_0^2 - 2 \vec{Q}_L ^2} 
\end{equation}
Finally we calculate the Euler number of this solution for $\beta \to 0$.
For this calculation, we use (\ref{vol}) directly 
because (\ref{diag}) is the equation for the 
diagonal metric. The result is that 
\begin{equation}
 \chi = \frac{1}{32\pi}\int_M \epsilon_{abcd}R^{ab} \land R^{cd} = 1.
\end{equation}
For the extremal metric, we need to use both the volume term(\ref{vol}) 
and the boundary term(\ref{bound}) in order to get the integer number 
as the Euler number. 
In contrast, our solution only need the volume term(\ref{vol}), 
namely $\partial V \neq \partial M$.
Therefore we conclude that this solution is non-extremal. 

\section{Black Hole Entropy}

In this section we evaluate the entropy of the black hole solution
we have found in the previous section semiclassically.
We calculate the entropy of the black hole using the following method 
instead of 
the ordinary equation $S_{ent} = \frac{Area}{4}$. We use the metric 
\begin{eqnarray}
 ds^2 = e^{2U(r)}(dt^2 +m_0 \cos \theta d\phi)^2 + e^{-2U(r)}dr^2 
  + R(r)^2 d\Omega^2 
\end{eqnarray}

The vierbeins are

\begin{eqnarray*}
   && \hat{\theta}^0 = e^U (dt + m_0 \cos \theta d \phi), \\ 
   && \hat{\theta}^1 = e^{-U} dr,    \\ 
   && \hat{\theta}^{2} = R d \theta, \\
   && \hat{\theta}^{3} = R \sin \theta d \phi.
\end{eqnarray*}

Then the entropy of the black hole is written as~\cite{GK,LP,KOP} :
\begin{eqnarray}
 S_{ent} &=& - \frac{1}{8\pi}(\int_{\partial V} - \int_{\partial M})
           [K] d\Sigma 
           = - \frac{1}{8\pi}(\int_{\partial V} - \int_{\partial M})
           [K]
         \hat{\theta}^{0} \land \hat{\theta}^{2} \land \hat{\theta}^{3}
          \nonumber \\ 
      &=& \frac{1}{8\pi} \frac{1}{\sqrt{g_{rr}}} 
          \frac{\partial}{\partial r}
          \int_{\partial M} 
          \hat{\theta}^t \land \hat{\theta}^{\theta} \land \hat{\theta}^{\phi}
           \bigg|_{r=r_H}    
            \nonumber \\
      &=& \frac{\beta R}{2}(R \partial_r U + 2 \partial_r R) e^{2U}
          \bigg|_{r=r_H} \nonumber \\ 
      &=& \frac{Area}{4} + \frac{\beta R}{2}
           [e^{2U} 2 \partial _rR ] \bigg|_{r=r_H} \label{taub} 
\end{eqnarray}
Here the second term in the last line vanishes when we consider the 
solution like the Schwarzschild or RN solutions. 

It is not clear whether the explicit relation between the entropy and the 
Euler number exists for the Taub-NUT type metric 
unlike the spherically symmetric solutions which are discussed in the 
section 2. However the important point is that 
the entropy inevitably vanishes when $\partial V = \partial M$ 
in (\ref{taub}). 
For our solution, $\partial V$ does not coincide $\partial M$ 
as it is discussed in the section 3.
Therefore this solution is non-extremal and it can possess 
the non-zero entropy.

In contrast, for the extremal black holes 
it is necessary that 
$\partial V$ must coincide with $\partial M$ 
because of their Euler numbers 
as is discussed in section 2. 
Therefore their entropies are zero 
even if their areas of the event horizons are non-vanishing.

For our solution, we calculate the entropy of black hole by this formula. 
Its area vanishes because $r_H = 0$. However this is not the essential 
problem unlike 
the vanishing Euler number. In order to 
get a non-zero area, we introduce the stretched horizon~\cite{STU} as 

\begin{equation}
  r_H = C :constant. 
\end{equation}
In contrast, it is not enough just to introduce the stretched horizon if we 
consider the solutions whose Euler numbers are zero. 
If we introduce stretched horizons to both $\partial V$ and $\partial
M$, the entropy still vanishes.
We believe that the stringy effects 
can be neglected because we study the BPS saturated states. 
Therefore it is possible to use our solution in order to discuss the 
relation between 
the entropy and degeneracy of the string states. 

We clarify the difference between the Sen's solution and our solution. 
The Sen's solution is extremal because its Euler number is zero (as we 
evaluated it in the previous section). 
According to Hawking's argument, Sen's solution has no entropy. On the 
other hand our solution is not extremal 
because its Euler number is non-zero. Hence it can possess non-zero entropy. 
Therefore we think that the entropy of our solution can be understood as 
the microscopic entropy 
which counts the elementary string states without any complications. 

In this way we obtain the entropy of our solution :

\begin{equation}
 S_{ent} = 3 \pi C n_0 = 3 \pi C \sqrt{m_0^2 - 2 \vec{Q}_L^2} 
\end{equation}
We compare this entropy with the statistical entropy which counts the 
degeneracy of elementary string states. 
The mass formula for heterotic string is: 
\begin{equation}
 m_0^2 = \frac{\vec{Q}^2_R}{8g^2} 
 = \frac{g^2}{8} (\frac{\vec{Q}^2_L}{g^4} + 2 N_L - 2) 
\quad (N_R = \frac{1}{2}) 
\end{equation}
where $N_L$ is the total oscillator contribution to the squared mass from 
the left moving oscillator. 
$g = <e^{\frac{\phi}{2}}>_{\infty}$ is the vacuum expectation value of the 
dilaton field at infinity. 
The degeneracy of the states is given by
\begin{equation}
  d_{E.S} \simeq \exp (4\pi \sqrt{N_L}).
\end{equation}
Thus, the entropy, which is calculated from the elementary string spectrum, 
is given by 
\begin{equation}
 S_{E.S} \equiv \ln d_{E.S} \simeq 4\pi\sqrt{N_L} 
 \simeq \frac{8\pi}{g}\sqrt{m_0^2 - \frac{\vec{Q}_L^2}{8g^2}}.
\end{equation}
If we make a choice that $g=\frac{1}{4}$ and $C=\frac{32}{3}$, then we 
obtain that $S_{E.S} = S_{ent}$. 

\section{Conclusion} 

We have derived the new solution of the effective heterotic string action, 
which is non-extremal and is BPS saturated. 
This solution has non-zero entropy because its Euler number is non-zero. 
That is consistent with the Hawking's argument that the extremal black 
holes don't have the entropy. Furthermore this solution 
can be treated semiclassically because it is the BPS saturated state. 
Consequently we can interpret the entropy of this black hole 
microscopically by the degeneracy of string states as Sen did. 
In comparison to Sen's solution which is extremal,
our solution is non-extremal and it can possess non-vanishing entropy.
Therefore the entropy of our solution can be understood as the microscopic 
entropy which counts the elementary string states 
in a straightforward way. 
We expect that we can find other interesting solutions of string theory 
from the classical solution of the Einstein equation in an 
analogous method. 

\vspace{0.5cm}

{\Large{\bf Acknowledgements}} 

\vspace{.5cm}

We thank Y.Kitazawa for discussions and for carefully reading the manu-script 
and suggesting various improvements.

\end{document}